\documentclass[journal,comsoc, 12pt]{IEEEtran}
\usepackage{amsmath,amsfonts}
\usepackage{algorithmic}
\usepackage{algorithm}
\usepackage{array}
\usepackage[caption=false,font=normalsize,labelfont=sf,textfont=sf]{subfig}
\usepackage{textcomp}
\usepackage{stfloats}
\usepackage{url}
\usepackage{verbatim}
\usepackage{graphicx}
\usepackage{cite}
\begin{document}

\title{Quantum Wrapper Networking}

\author{S. J. Ben Yoo,~\IEEEmembership{Fellow,~IEEE}, Sandeep Kumar Singh, Mehmet Berkay On, \\ Gamze G\"ul,   Gregory S. Kanter, Roberto Proietti, and Prem Kumar~\IEEEmembership{Fellow,~IEEE}
\thanks{S. J. B. Yoo, S. K. Singh, and M. B. On are with the Department of Electrical and Computer Engineering, University of California, Davis, Davis, CA 95616, USA (Email: sbyoo@ucdavis.edu). \\
R. Proietti is with the Department of Electronics and Telecommunications Engineering, Politecnico di Torino, Italy. \\ G. G\"ul, G. S. Kanter, and P. Kumar are with the Center for Photonic Communication and Computing, Department of Electrical and Computer Engineering, Northwestern University, Evanston, IL 60208, USA (Email: kumarp@northwestern.edu).
}
}



\maketitle

\begin{abstract}
We introduce a new concept of Quantum Wrapper Networking, which enables control, management, and operation of quantum networks that can co-exist with classical networks while keeping the requirements for quantum networks intact. The quantum wrapper networks (QWNs) enable the transparent and interoperable transportation of quantum wrapper datagrams consisting of quantum payloads and, notably, classical headers to facilitate the datagram switching without measuring or disturbing the qubits of the quantum payload. Furthermore, QWNs can utilize the common network control and management for performance monitoring on the classical header and infer the  quantum channel quality. 
\end{abstract}

\begin{IEEEkeywords}
quantum networks, entanglement, optical label switching, transparent optical networks.
\end{IEEEkeywords}

\section{Introduction}
\IEEEPARstart{T}{he} realization and deployment of a large-scale quantum network will truly revolutionize the way we conduct our lives \cite{Simon2017TowardsNetwork}. The prospect of a quantum internet \cite{Kimble2008TheInternet,Lloyd2004InfrastructureInternet,Wehner2018QuantumAhead} with quantum entanglement between any two points on the Earth's surface is full of thrilling possibilities. Recent progress in quantum technologies, especially quantum sources, repeaters, processors, memories, detectors, etc., is encouraging \cite{Simon2017TowardsNetwork,Kimble2008TheInternet,Lloyd2004InfrastructureInternet,Wehner2018QuantumAhead}. When quantum computers become available and commercially viable, then they will be able to solve problems that are exponentially difficult for classical computers. The next question is how to interconnect quantum computers for communication and distributed computing. 

A quantum network could be built on existing fiber-optic infrastructure, as has been demonstrated for quantum key distribution and entanglement distribution networks \cite{joshi2020trusted,pant2019routing,pompili2022experimental}, with low-loss classical and quantum devices as well as communication protocols. 
However, despite some progress in developing quantum computing, networking, and related device technologies, the quantum internet is far from reality.  
Qubits are extremely fragile, and their fundamental principles forbid measurements or monitoring required for network control and management. However, the fragile qubits need to be transported and switched while guaranteeing their integrity with some level of Quality of Transmission (QoT) and Quality of Entanglement (QoE) assurances. Furthermore, if and when such a quantum networking technology becomes available, our history tells us that successful transitions in networking technologies would require a strategy for seamless upgrades from today’s classical networks to the new quantum networks with interoperability.  In considering the development and deployment of quantum networks, many new challenging questions arise:  
\begin{itemize}
    \item How do you place a control plane on quantum networks?
    \item How do you manage quantum networks?
    \item How do you monitor the performance of quantum networks?
    \item How do you achieve transport in quantum networks?
    \item How do you switch and route in quantum networks?
    \item Do we have simulation tools and experimental testbeds to design, simulate, and operate quantum networks with quantum devices before actual deployment?
\end{itemize}

Again, if we somehow manage to find solutions to all of the above, we cannot envision dismantling and overhauling today’s classical networks to place a new quantum network.  Therefore, in order to enable upgrades from the current network to the future quantum internet, we need a quantum network that can be deployed independently or with the current classical network and be backward compatible.

\section{Quantum Wrapper Networking Architecture And Protocols}
\begin{figure*}[!t]
\centering
\includegraphics[width=1\linewidth]{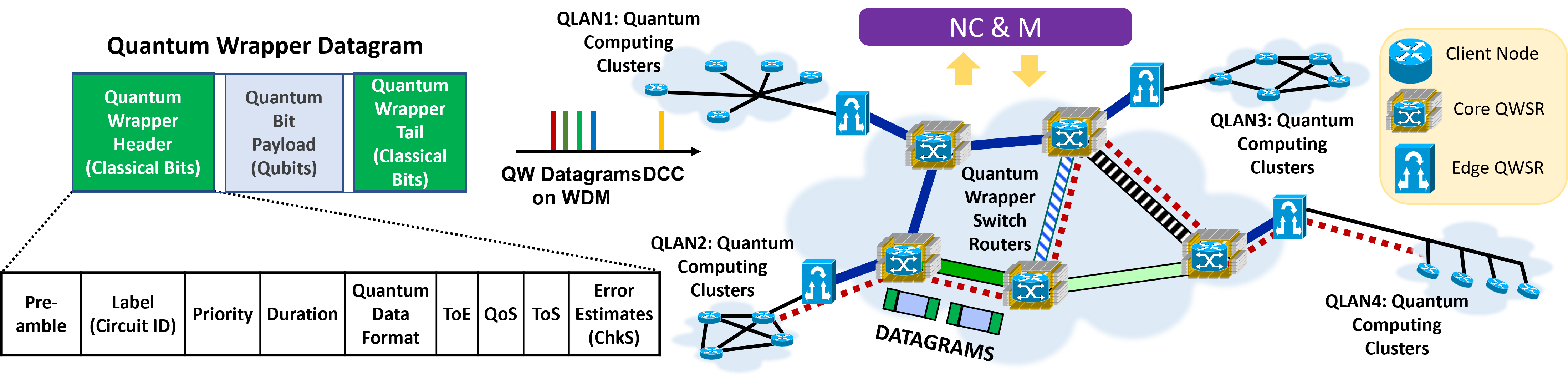}
\caption{Quantum networking with Quantum Wrappers (QWs).  A quantum wrapper datagram consists of a Quantum Wrapper Header (classical bits), a Quantum Wrapper Payload (quantum bits), and a Quantum Wrapper Tail.  Quantum Wrapper Switches will forward the datagrams, establish data flows, setup/tear down circuits utilizing the quantum wrapper headers and data communication channel (DCC).  The quantum edge nodes will interface with quantum network clients and add Quantum Wrappers to the quantum payloads accordingly.  The Quantum Wrapper Switches include Quantum Wrapper swappers which will replace the old Quantum Wrapper with a new Quantum Wrapper while keeping the same qubit payload.  The Quantum Wrapper Header consists of a preamble, circuit ID (or source-destination), priority, duration of the qubit payload, type of entanglement, quality of service, type of service (e.g., real or non-real time application), and checksum bits.}
\label{fig:qwn_arch}
\end{figure*}

We propose a new networking technology called ‘\emph{Quantum Wrapper}’ (QW) to enable quantum networking on today’s networking platform. As Fig.~\ref{fig:qwn_arch} illustrates, the quantum wrapper consists of the QW Header and the QW Tail in the form of classical bits containing information pertaining to routing, multiplexing, timing, format (e.g., type of entanglement, and wavelength), etc. These classical bits will ‘wrap’ the quantum payload (qubits)\footnote{Although in this paper we focus on qubits as the quantum payload, i.e., discrete-variable quantum information, the ideas developed are equally applicable to qmodes as the quantum payload, i.e., continuous-variable quantum information.} to facilitate end-to-end transport of the quantum payload without having to read or alter the quantum data.  

As in the Optical Label Switching (OLS) networking technology \cite{Yoo2006OpticalInternet}, where data payload of any format and protocol can be transported and switched utilizing the optical label of a standard format (e.g., 40 bits at 1\,Gb/s rate) to offer packet, burst, and circuit switching, the QW will utilize a predetermined format for the Wrapper bits while the quantum payload can assume any qubit format (such as polarization, time-bin, or frequency-bin encoding) and protocol (discrete variable (DV) or continuous variable (CV)). While the QW can adopt a very similar overhead format as in Optical Transport Network (OTN), it is conceptually meaningful to illustrate one of the QW header formats in Fig.~\ref{fig:qwn_arch}.  The Circuit ID on the header can be source-destination pair, MPLS (Multi-Protocol Label Switching)-like label, or a circuit ID.  QWs will allow MPLS-like label switching and flow switching of datagrams as well as traffic engineering.  As we expect that qubit payloads will be of variable size (from  hundreds of kilo-qubits to  giga-qubits) and bursty, we reserve Class C and Class D QW for future optical packet switching.   As Fig.~\ref{fig:qwn_arch} shows, the QW datagrams can be on wavelength-division multiplexed (WDM) data channels, while a data communication channel (DCC) on a separate wavelength can support the control plane and management plane communications for Software Defined Networking (SDN).  
As we will discuss further in more detail, QW networking has the following key approaches to address the challenging questions mentioned earlier.
\begin{itemize}
    \item QW networking uses Classical Quantum Wrappers for transport, switching, and routing in the quantum network without reading the Quantum Data Payload (qubits) until the qubit receiver.
    \item QW networking uses classical bits in the Wrapper to conduct optical performance monitoring to         infer Signal-to-Noise-Ratio (SNR), QoT, Dispersion, Polarization Mode Dispersion (PMD), etc., without touching the qubits.
    \item QW networking supports data payload of any protocol and format. 
    \item QW networking achieves full compatibility with Software Defined Networking (SDN).
    \item QW networking achieves interoperability and backward-compatibility with existing or legacy  telecom protocols: Ethernet, OTN, MPLS, etc.
    \item QW networking requires no strict synchronization of the network. QW can be used for timing synchronization or polarization stabilization.
    \item QW networking exploits much of the existing control plane protocols to allow backward compatibility and seamless upgrades from today’s networks to the future quantum internet.
\end{itemize}

\begin{figure*}[!t]
\centering
\includegraphics[width=1\linewidth]{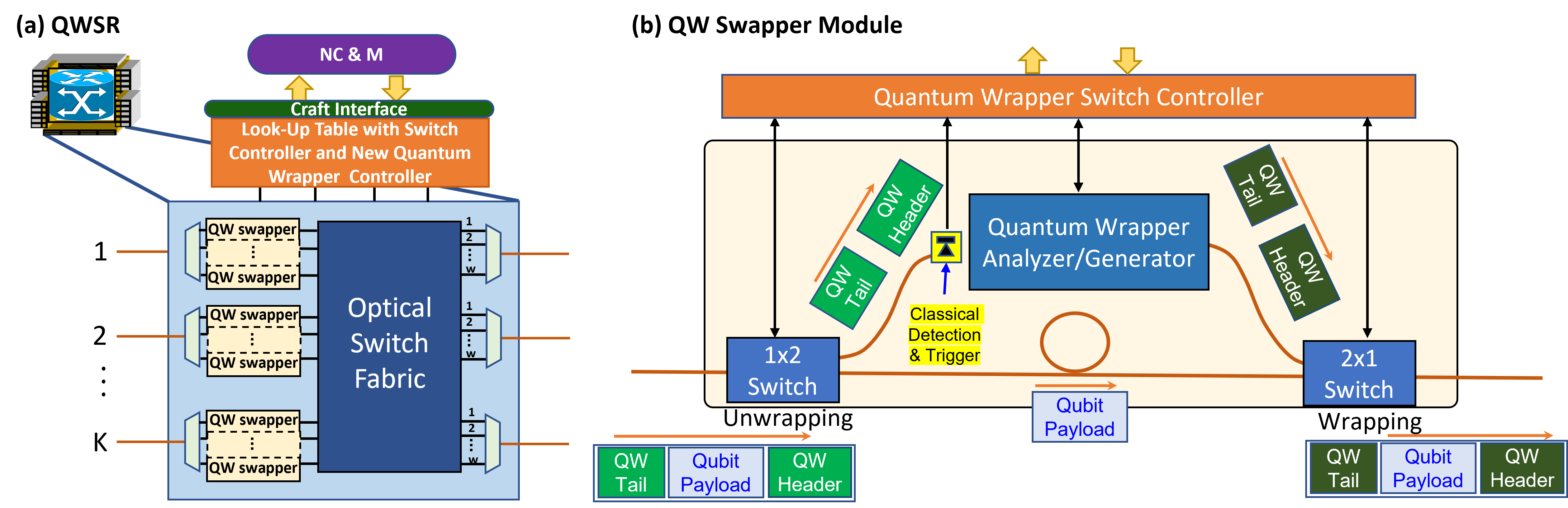}
\caption{(a) Quantum Wrapper Switching Router consists of QW swapping modules with Optical Switch fabric.  (b) A QW swapping module consists of a lookup table with controllers that interface with NC\&M, 1x2 and 2x1 switches, fiber delay line(s), and a QW generator. A $1\times2$ switch unwraps the header and tail for classical detection and triggers for switching and new wrapper generation. Meanwhile, the polarization-entangled Qubits payload is delayed. Again, a $2\times1$ switch wraps the Qubit payload with a new QW header and tail and directs the new datagram to the Optical Switching Fabric.}
\label{fig:qw_switch}
\end{figure*}

\subsection{Quantum Wrapper Switching Routers (QWSRs)}
As Fig.~\ref{fig:qw_switch}(a) illustrates, QWSRs in Fig.~\ref{fig:qwn_arch} may consist of a lookup table with controllers that interface with network control and management (NC\&M), optical switch fabric, and QW swapping modules each of which consists of $1\times2$ switches, fiber delay line(s), and a QW generator.  Figure~\ref{fig:qw_switch}(b) shows an example of a time-multiplexed packetized transmission of the QW header and tail with polarization-entangled qubit payload. The QW swapping mechanism of the qubit payload is also shown.  

\subsection{Control Plane for QW Networks}
The out-of-band DCC provides communications between the Quantum Nodes to form a Data Communication Network (DCN) for a centralized Network Control and Management (NC\&M) while the Quantum Wrapper offers in-band signaling for the datagram with an opportunity for distributed control planes at each Quantum Node.  The in-band signaling dictates the rapid and local responses, including the forwarding of the datagram based on the forwarding table and the label content. The DCN allows communications between the network elements, updates the forwarding table, and allows programming of the programmable network elements based on the network traffic conditions.
This also interoperates with the NC\&M system to achieve QW label distribution and to update the forwarding table of routers and switches at the edge and core of the network. 

\subsection{Monitoring in QW Networks}
Networks require a) a service plane, b) a management plane, c) a control plane, and d) a data plane. The management plane governs the devices in the network and the state of those devices, e.g., power, configuration, operational readiness, etc.  The management plane has five main roles \cite{Wehner2018QuantumAhead}: a) performance management, b)  fault management, c) configuration management, d) security management, and e) accounting management.  Monitoring is an essential function for the proper management and operation of any network.  In the classical hardware-defined electronic networks, nearly all monitoring took place in the electronics, and it came for free since linecards or interfaces to network elements (electronic switches, routers, terminals, etc.) were equipped with such monitoring functions at layer 1, layer 2, layer 3, etc. that looked into signal integrity, bit errors, framing errors, address errors, etc. that in turn fed such information to the management plane for conducting the five roles mentioned above.  

In a classical optical software-defined network (SDN), the optical monitoring functions become very difficult if transparent end-to-end connections are desired.  For the proposed quantum optical networks with SDN control planes, we expect to find even more significant challenges due to the quantum nature of the qubits.  One solution to this challenge is to employ the Quantum Wrapper protocol discussed above.  
Monitoring the classical bits in the framing bits of a Quantum Wrapper can allow inferred monitoring of the qubit payload.  For instance, the framing bits can include source, destination, traffic engineering, check-sum, etc., and offer information pertaining to \emph{what type of packets are going from where to where with what type of service (ToS), class-of-service (CoS), and quality-of-service (QoS)}, and detecting these framing bits electrically and optically will offer optical wavelength monitoring, optical signal-to-noise monitoring, and bit-error-rate (BER) estimations.

The Quantum Wrapper based quantum networking is especially exciting not only because it offers a rich set of attribute-based networking encoded on the wrapper frame bits (labels), but because it also offers a means to conduct real-time optical performance monitoring.  This allows maintaining the required quality of service (QoS) and high level of performance even for signals that experience time-varying physical layer impairments (QoS-aware and impairment-responsive networking) \cite{Geisler2011ExperimentalAwareness}. Here, we can encode supervisory channel information at low speed (e.g., 1.25\,Gb/s) in the Wrapper that precedes in time relative to the qubit data payload. Since the data and the supervisory channel signal follow that same path, many properties of the quantum and classical channels would be highly correlated, including loss, PMD (under time-multiplexing of header and payload on the same wavelength) or chromatic dispersion (under time-wavelength multiplexing), and polarization-dependent loss. This affords the opportunity to use the classical channel to monitor such properties, giving insight into the quantum channel integrity. Thus, by incorporating the supervisory channel quantum BER (qBER) measurement in the control plane distributed over the network, one can map the link state condition across the network and adjust the physical layer parameters. 

There are, however,  a large number of things that can affect the fidelity of the qubits that would not show up in the supervisory channel. For example, $10^{-2}$ noise photons/ns will not show up in the classical channel but destroy most of the quantum channels.
Nevertheless, the ability to periodically send an empty ``supervisory" packet containing attenuated coherent-state light would allow the actual quantum equipment (single photon detector, quantum memory, etc.) to make quality measurements pertinent to the qubit system. This kind of measurement can reveal details of the quantum channel and the quantum component performance relatively quickly and in a controlled way.

\begin{figure*}[!t]
\centering
\includegraphics[width=0.95\linewidth]{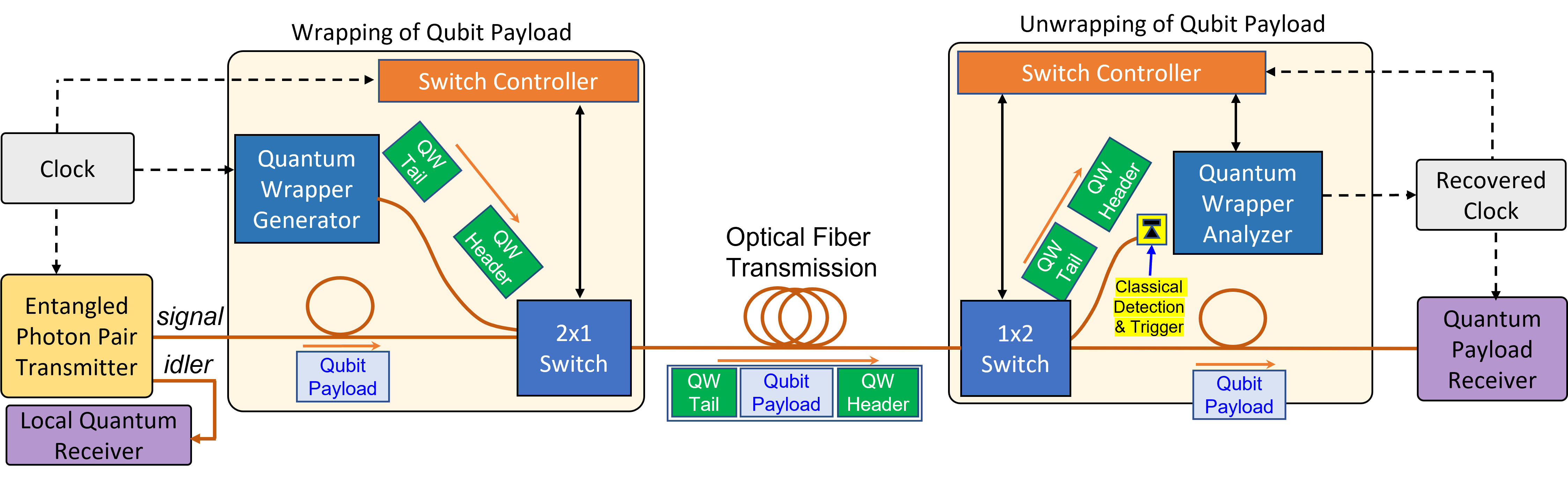}
\caption{A schematic of the transmission of quantum wrapper datagrams on a point-to-point optical fiber link. An Entangled Photon Pair Transmitter generates \emph{signal} and \emph{idler} pair photons. A Qubit payload consists of polarization-entangled \emph{signal} photons generated around 1565.72\,nm. A Quantum Wrapper Generator generates QW Header and Tail around 1561.42\,nm. QW wraps the payload with the help of a $2\times 1$ Switch controlled by a synchronized clock. At the reception, a $2\times 1$ Switch unwraps QW header and tail to analyze them together with the received Qubit payload with the help of a recovered clock.}
\label{exp_setup}
\end{figure*}
\vspace{-0.1cm}


\subsection{TCP/IP for QW Networks}
TCP/IP (Transport Control Protocol over Internet Protocol) of the quantum networks can also be achieved using the quantum wrapper by implementing windowing and acknowledgment (ACK) or negative acknowledgment (NACK) functions when the QW header has been received (or timed out).  Again, this is accomplished without having to read the qubit payload. 

\subsection{Quantum Entanglement Distribution}
One type of quantum payload would be packets of entangled qubits to be distributed to pairs of quantum nodes on the network for the purpose of quantum teleportation between the node pairs. Our quantum wrapper technology could be utilized to ensure that entanglement is distributed to the requesting node pairs at the highest entanglement fidelity possible. 


\subsection{Quantum Repeaters}
Quantum repeaters are analogous to optical amplifiers in classical transmission. Since qubits cannot be duplicated (amplified), an optical amplifier in the quantum domain is not allowable for photonic qubits carrying quantum information across nodes in a network. On the other hand, quantum repeater concepts have been devised, and significant progress has been made recently \cite{Muralidharan2016OptimalCommunication,Bhaskar2020ExperimentalCommunication}. The proposed quantum wrapper technology facilitates the incorporation of the quantum repeaters in the quantum core nodes by the inclusion of the QW switch reader.

\section{Preliminary Experiments For Quantum Wrapper Networking}
We performed some preliminary experiments to gain insight into the feasibility of QW networking~\cite{mehmet2023ofc}. A schematic for transmitting packetized classical headers and gated quantum payloads by wavelength-time multiplexing is shown in Fig.~\ref{exp_setup}. 
For packetized transmission, we generated burst-mode headers by gating the continuous classical data stream generated by a transmitter (TX) with  an external Mach-Zehnder modulator (MZM). On the other hand, the quantum payloads were realized by gating the single-photon detectors (SPD) during their occurrence with a synchronous electrical signal.  
We used a 1.23\,ms QW datagram period consisting of 1.13\,ms quantum payload and 102.4\,$\mu$s QW header. In the absence of a low-loss nanosecond $2\times1$ switch, we used a 2-by-2 power coupler to combine Header bits with the quantum signal on separate wavelengths and transmit them together over an optical fiber.  

\begin{figure}[htbp]
  \centering
  \includegraphics[width=0.4\textwidth,height=\textheight,keepaspectratio]{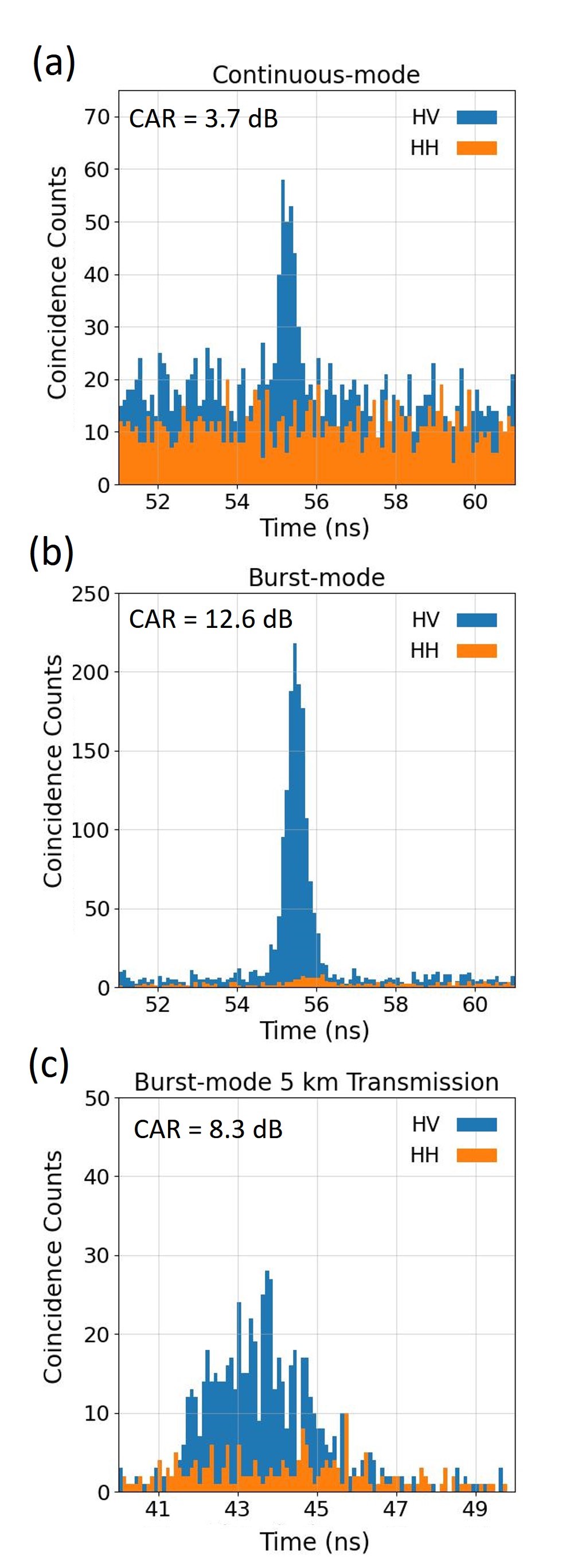}
\caption{(a) Coincidence counts with WDM continuos-mode headers.  (b) Coincidence counts with WDM burst-mode headers. (c) Coincidence counts with WDM burst-mode headers with 5 km transmission. Classical Transmitter launch power is -21\,dBm \cite{mehmet2023ofc}.}
\label{results-cc}
\end{figure}

\begin{figure}[htbp]
  \centering
  \includegraphics[width=0.5\textwidth,height=\textheight,keepaspectratio]{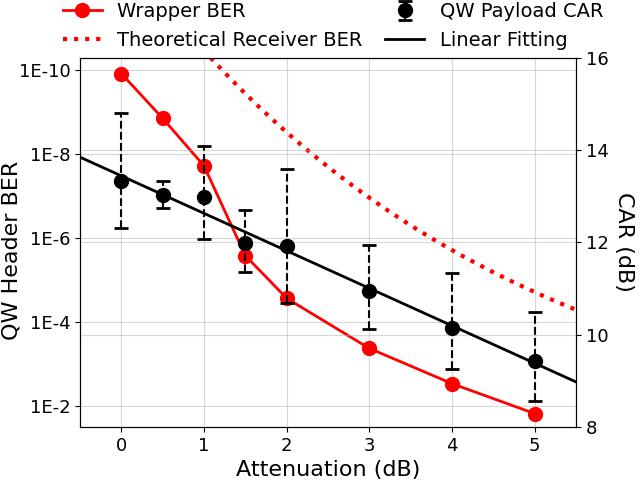}
\caption{Header BER and qubit CAR vs. channel attenuation at -28\,dBm TX launch power. \cite{mehmet2023ofc}.}
\label{results-monitor}
\end{figure}

To realize the Qubit payload transmitter, we used a polarization-entangled photon-pair source that generates entangled photon-pairs, \textit{signal} in the L-band and \textit{idler} in the C-band.  A variable optical attenuator (VOA) in the coexistence channel emulates different possible values of the coexistence channel attenuation. A wavelength-division multiplexing (WDM) device demultiplexes classical and quantum signals. The demultiplexed classical signal is preamplified and detected by an optical receiver. 
Fiber-based polarization controllers (FPC) are aligned to the horizontal (H) and vertical (V) polarization axes of the polarization beam splitters (PBSs).

We recorded coincidence counts (CC) in 2\,ns time windows for both continuous and burst-mode header bits at TX launch power of $-21$\,dBm. Header BER is $<10^{-12}$ at this power level. We computed the coincidence-to-accidental ratio (CAR), which represents the quality of the quantum channel, as the ratio of overall coincidence counts and the accidental counts due to system noise. Figure~\ref{results-cc} shows coincidence counts in 100 seconds for continuous and burst modes.  We observe that the
CAR of the qubits degrades significantly from 16 dB without a header to 3.7\,dB with the addition of a continuous header due in part to the $-50$\,dB crosstalk of the demultiplexing WDM device. However, by time-division multiplexing the header, the CAR recovers to 12.6\,dB. A higher extinction ratio MZM could further improve the CAR with the burst-mode operation. We also performed relatively long-distance copropagation of quantum payloads with classical headers using a 5\,km SMF-28 fiber in the coexistence channel and measured the coincidence counts in the burst mode. The coincidence count distribution spreads out in time due to the interplay between the large 40\,nm quantum signal bandwidth and the dispersion of the link (see Fig.~\ref{results-cc} (c)). Further investigation is needed to better understand and account for fiber transmission impairments.  

When the TX launch power is above $-28$\,dBm, we observed error-free transmission for the burst header. Thus, we increased the coexistence channel attenuation to obtain a  relation between the QW header BER and the QW payload CAR (see Fig.~\ref{results-monitor}). Obviously, increasing the attenuation in the channel degrades the QW header's SNR and causes higher BER. Similarly, the CAR as \textit{Quantum SNR} decreases with increasing channel attenuation. Therefore, in this case, the QW header's BER can infer quantum channel quality without directly measuring the qubits. 

\section{Conclusions and Future Directions}
We introduced a new concept of Quantum Wrapper Networking where quantum networks can co-exist with classical networks utilizing the common network control and management and performance monitoring frameworks while keeping the requirements for quantum networks intact. Through some preliminary experiments, we demonstrated packetized transmission of the quantum wrapper datagrams in a point-to-point link and observed that it is possible to monitor the quantum channel quality without measuring qubits with the help of quantum wrapper bits. 

As this article highlighted, there are many challenges in realizing quantum networks---from the application layer to the physical layer. Thus, we need to come up with innovative solutions. Especially in quantum wrapper networks, in addition to the development of a software-control and management, switching the quantum payloads when a specific quantum wrapper header is detected is a key. Although using quantum wrapper headers to monitor quantum channels is interesting, the qubit entanglement fidelity depends on many factors, including quantum noise, which will not be detected by the quantum wrapper bits. Furthermore, developing timing synchronization and polarization stabilization techniques using quantum wrapper bits would facilitate the deployment of quantum wrapper networks paving the way toward the future quantum internet.

\section*{Acknowledgement}
This work is supported by the U.S. DOE, Office of ASCR program under Award Number DE-SC-0022336.

\bibliographystyle{IEEEtran}

\bibliography{mendeley_references,ref_local}

\section{Biography}
\vspace{-1cm}
\begin{IEEEbiographynophoto}{S. J. Ben Yoo} [S’82, M’84, SM’97, F’07] received the B.S., M.S., and Ph.D. degrees from Stanford University, CA, USA, in 1984,
1986, and 1991, respectively. He is a Distinguished Professor of electrical
engineering at UC Davis. His research at UC Davis includes future computing, photonic communications, cognitive networking, and integrated systems for the future Internet. He is a recipient of the DARPA Award for Sustained Excellence (1997), the Bellcore CEO Award (1998), the Mid-Career Research
Faculty Award (2004 UC Davis), and the Senior Research Faculty Award
(2011 UC Davis).
\end{IEEEbiographynophoto}
\vspace{-1cm}
\begin{IEEEbiographynophoto}{Sandeep Kumar Singh} received M.S. from Indian Institute of Technology Madras, India, and  Ph.D. from Technical University of Braunschweig, Germany in 2014 and 2019. He was a research scientist at the German Aerospace Center from 2019 to 2021. He is currently a postdoctoral scholar at UC Davis. His areas of interest include quantum networking and data center networking.
\end{IEEEbiographynophoto}
\vspace{-1cm}
\begin{IEEEbiographynophoto}{Mehmet Berkay On}
received the B.S. from Bilkent University, Ankara, Turkey in 2018. He is currently working towards a Ph.D. degree at UC Davis. His research interests are photonic neuromorphic computing, quantum networks, and compressive sensing.
\end{IEEEbiographynophoto}
\vspace{-1cm}
\begin{IEEEbiographynophoto}{Gamze G\"ul}
received B.S. from Bilkent University, Ankara, Turkey, and M.S. from Koc University, Istanbul, Turkey in 2017 and 2019. She is a Ph.D. student at Northwestern University. Her research interests are quantum communications and integrated photonics.
\end{IEEEbiographynophoto}
\vspace{-1cm}
\begin{IEEEbiographynophoto}{Gregory S. Kanter} received M.S., Ph.D., and MEM degrees from Northwestern University in Evanston, Il. USA, in 1996, 2000, and 2009. He is a Research Associate Professor at Northwestern University and CEO of NuCrypt, LLC. His research includes quantum networking, nonlinear fiber optics, optical communications, and RF photonics, and he holds over a dozen related patents. 
\end{IEEEbiographynophoto}
\vspace{-1cm}
\begin{IEEEbiographynophoto}{Roberto Proietti} received his M.S. degree from the University of Pisa in 2004, and his Ph.D. degree from Scuola Superiore Sant Anna in 2009. He was a project scientist at UC Davis. He is currently an assistant professor with Politecnico di Torino. His research interests include optical switching technologies, architectures for HPC/data centers, and quantum networking.
\end{IEEEbiographynophoto}
\vspace{-1cm}
\begin{IEEEbiographynophoto}{Prem Kumar}
[M’89, F’03, LF’23] is a Professor of Information Technology at Northwestern University with a primary appointment in Electrical and Computer Engineering. His current research focuses on quantum communications and networking for interconnecting future quantum computing and sensing systems. He received his Ph.D. in 1980 from SUNY/Buffalo and joined Northwestern in 1986 after spending 5 years at MIT. He has served as a Program Manager at DARPA (PM of the year, 2015; Secretary of Defense Medal for Outstanding Public Service, 2016). He has been a Distinguished Lecturer for the IEEE Photonics Society (2008-2010), Hermann Haus Lecturer at MIT (2013), recipient of the Quantum Communication Award from Tamagawa University in Tokyo, Japan (2004), and the Walder Research Excellence Award from the Provost’s office at Northwestern (2006).
\end{IEEEbiographynophoto}

\vfill

\end{document}